\documentclass[conference,10pt]{IEEEtran}
\usepackage[dvips]{graphicx}
\usepackage{amsmath,amssymb,color}
\usepackage{multirow}

\definecolor{nblue}{rgb}{0,0,0.6}
\definecolor{nred}{rgb}{0.8,0,0}
\definecolor{ngreen}{rgb}{0,0.4,0}
\definecolor{norange}{rgb}{1.0,0.5,0.3}

\definecolor{scblue}{rgb}{0,0,0.4}

\newenvironment{Enumerate}{\begin{enumerate}\setlength{\parsep}{0mm}\setlength{\parskip}{0mm}\setlength{\itemsep}{0.5ex}}{\end{enumerate}}
\newenvironment{Itemize}{\begin{itemize}\setlength{\parsep}{0mm}\setlength{\parskip}{0mm}\setlength{\itemsep}{0.5ex}}{\end{itemize}}

\newenvironment{MValue}{\left\{ \begin{array}{ll}}{\end{array} \right.}

\newcommand{\Bf}[1]{{\bf #1}}
\newcommand{\BM}[1]{\mathbf{#1}}
\newcommand{\BS}[1]{\boldsymbol{#1}}

\newcommand{\MR}[1]{\mathrm{#1}}

\newcommand{\Funcf}[3]{#1\!\left(\frac{#2}{#3}\right)}

\newcommand{\Brace}[1]{\left\{ #1 \right\}}
\newcommand{\Brack}[1]{\left[ #1 \right]}
\newcommand{\Parth}[1]{\left( #1 \right)}

\newcommand{\Half}{\frac{1}{2}}

\newcommand{\RealS}{\mathbb{R}}

\newcommand{\Trns}[1]{{\bf #1}^{\mbox{\sf \tiny T}}}

\newcommand{\ON}{{\cal O}(N)}

\begin{document}

\title{Highly Efficient Non-Separable Transforms for Next Generation Video Coding}

\author{\IEEEauthorblockN{Amir Said, Xin Zhao, Marta Karczewicz, Hilmi E. Egilmez, Vadim Seregin, Jianle Chen}
\IEEEauthorblockA{Qualcomm Technologies, Inc., San Diego, CA, USA}
}

\maketitle

\begin{abstract}
For the last few decades, the application of signal-adaptive transform coding to video compression has been stymied by the large computational complexity of matrix-based solutions. In this paper, we propose a novel parametric approach to greatly reduce the complexity without degrading the compression performance. In our approach, instead of following the conventional technique of identifying full transform matrices that yield best compression efficiency, we look for the best transform parameters defining a new class of transforms, called HyGTs, which have low complexity implementations that are easy to parallelize. The proposed HyGTs are implemented as an extension of High Efficiency Video Coding (HEVC), and our comprehensive experimental results demonstrate that proposed HyGTs improve average coding gain by 6\% bit rate reduction, while using 6.8~times less memory than KLT matrices.
\end{abstract}


\section{Introduction}

Transform coding is a fundamental part in all the widely used video compression standards~\cite{Richardson:10:avc,Wien:15:HEV,Mukherjee:13:tlo}, which typically adopt a fixed, separable transform in order to maintain low complexity. For this purpose, the discrete cosine transform (DCT) is by far the most popular transform in such standards, since it can achieve good compression rates especially when inter-sample correlations between pixels are large, and it has fast implementations. However, the DCT is optimal under very restrictive assumptions, which do not practically hold for the most image and video data. On the other hand, it has been long known that the signal-adaptive Karhunen-Lo\`{e}ve Transform\footnote{What is discussed here applies to all forms of adaptive transform coding, but to avoid confusion we only refer to the KLT.} (KLT) is optimal for transform coding~\cite{Sayood:00:idc,Taubman:02:jsi,Woods:11:msp,Pearlman:11:dsc}. There is a good number of studies showing that, when combined with proper classification and adaptation methods, KLTs can very effectively improve compression efficiency. Nevertheless, since they used the inefficient matrix-based implementations, those contributions have not been adopted, nor further developed.

\begin{figure}
\centering
\includegraphics[width=56mm]{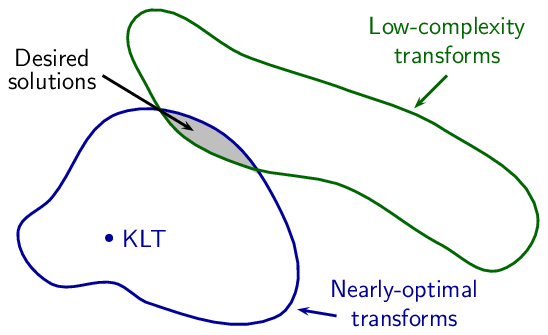}
\caption{\label{fg:GoodFast}The new design method exploits the fact that for media coding there is commonly a large set of nearly optimal transforms, and looks for those that that can also be computed with low complexity.}
\end{figure}

In this work, we propose a novel method that allows us to design more practical signal-adaptive transforms by taking simultaneously into account signal statistics, compression efficiency, and computational complexity. We consider the following basic premises about transform coding applied to media compression in our approach.
\begin{Itemize}
\item The mathematical tools to compute the optimal data-defined transform provide a unique type of solution. However, in practice transform coding gains are defined nearly exclusively by a few principal components (i.e., data within certain subspaces), and we commonly have large sets of transforms that produce coding results that are very close to the optimal.
\item There are certain orthogonal transform structures that enable computation with very low complexity (both memory and operations). Individually they cover a very limited range of all possible orthogonal transforms, but if they are properly combined, the range can be greatly extended, without significant increase in complexity.
\end{Itemize}
The novelty of our approach is that, instead of identifying the full transform matrices (e.g., KLTs) that yield best compression, we search for a set of parameters, which implicitly defines a \emph{low-complexity transform}, to derive a low-complexity transform that achieves the best coding performance. As illustrated in Fig.~\ref{fg:GoodFast}, we search for the intersection of two transform sets. The resulting transform should produce compression sufficiently close to the KLT's, but at much lower computational complexity. The low-complexity transforms are defined by a set of parameters which defines a special multi-pass decomposition of transform matrices, and they are computationally efficient and easily parallelizable by construction.  

In the rest of the paper, Section \ref{sc:Prelim} present some preliminaries about signal-adaptive transforms and computational complexity. In Section~\ref{sc:CMPT}, we introduce the general class of parametric multi-pass transforms (PMPT), and Hypercube-Givens transforms (HyGT) is proposed as a special case of PMPT in Section~\ref{sc:HyGT}. The HyGT parameter optimization is discussed in Section \ref{sc:HyGTOpt}. The experimental results and concluding remarks are presented in Sections \ref{sc:ExpRes} and \ref{sc:Concl}, respectively.

\section{Preliminaries}
\label{sc:Prelim}
\noindent{\textbf{{Signal-adaptive transform coding:}}} An essential aspect of advanced transform coding is that it should be highly adaptive, to exploit the non-stationary features of complex data. We assume that a vector $\Bf{x}$ with values of pixels in a video block is first subtracted from its prediction $\Bf{p}$ to create the zero-mean {\em residual vector} $\Bf{r}=\Bf{x}-\Bf{p}$, and that the decoder, using rate-distortion estimation, decides on a classification index $c\in\{1,2,\ldots,C\}$ for the current block, applies the transform corresponding to $c$ to obtain $\Bf{y}_c$, and encodes $c$ as side information.

To find the optimal transform for class $c$, we first compute (e.g., using training) the matrix with the conditional correlation of the residual vector
\begin{equation}
 \label{eq:CorrDefn}
 \BM{\Phi}_c = \MR{E}_{\Bf{r}|c} \Brace{ \Bf{r} \, \Trns{r} },
\end{equation}
which is symmetric and positive semi-definite, and we can always find one orthogonal matrix $\Bf{K}_c$ such that~\cite{Vaidyanathan:93:mfb,Strang:09:ila}
\begin{equation}
 \label{eq:Decorr}
 \Bf{K}_c \, \BM{\Phi}_c \, \Trns{K}_c = \BM{\Lambda}_c,
\end{equation}
where $\BM{\Lambda}_c$ is a diagonal matrix with the eigenvalues of $\BM{\Phi}_c$.

This {\em transform matrix} $\Bf{K}_c$ defines the KLT for $\BM{\Phi}_c$, and produces the transformed vector
\begin{equation}
 \label{eq:MatrxT}
 \Bf{y}_c = \Bf{K}_c \, \Bf{r},
\end{equation}
which has uncorrelated elements.

\noindent{\textbf{{Analysis of Computational Complexity}}}:
To analyze the computational complexity of transform coding, applied to images and video, we have to consider the following implementation choices: (a) Separable or non-separable; and (b) ``fast'' or matrix-based.

While any linear transformation can be implemented using a matrix multiplication as in eq.~(\ref{eq:MatrxT}), low-complexity computation has been available only for a few special ``fast'' transforms, like the DCT, which have very specific mathematical properties~\cite{Britanak:07:dxt}.

Table~\ref{tb:ExpBlkCmplx} shows the complexity (memory and operations) of transforms applied to blocks of $2^B \times 2^B$ residuals, considering that we use $C$ different transforms, according to their classification (the last row is explained in Section~\ref{sc:HyGT}). We can observe that the complexity of non-separable, matrix-based transforms---the ones that yield best compression---grows very fast, which is the reason they have been impractical except for very small blocks.

On the practical side, it is important to observe that while in the past arithmetic operations could have been much more expensive than memory, now we have to consider that the number of operations {\em per pixel} is not as critical, and instead we commonly need to limit the memory requirements, because best compression is obtained when the number of transforms $C$ is relatively large. 

\begin{table}
\centering
\caption{\label{tb:ExpBlkCmplx}Memory and arithmetic operations complexity of different types 2-D block transforms, applied to $2^B \times 2^B$ residual blocks, and considering $C$ different classifications.}
\begin{small}
\renewcommand{\arraystretch}{1.5}
\begin{tabular}{|l||c|c|c|} \hline
\bf Transform      & \bf Memory            & \bf Operations      & \bf Ops./pixel    \\ \hline \hline
 Fast, sep.        & O$\Parth{2^{B}  C B}$ & O$\Parth{2^{2B} B}$ & O$\Parth{B}$      \\ \hline
 Matrix, sep.      & O$\Parth{2^{2B} C}$   & O$\Parth{2^{3B}}$   & O$\Parth{2^B}$    \\ \hline
 Matrix, non-sep.  & O$\Parth{2^{4B} C}$   & O$\Parth{2^{4B}}$   & O$\Parth{2^{2B}}$ \\ \hline
 PMPT, non-sep.    & O$\Parth{2^{2B} C P}$ & O$\Parth{2^{2B} P}$ & O$\Parth{P}$      \\ \hline
\end{tabular} \end{small}
\end{table}

\section{Parametric Multi-pass Transform (PMPT)}\label{sc:CMPT}

In this section we discuss the basic approach for reducing the transform complexity, with technical details in the following sections. We start by considering that $N \times N$ orthogonal matrices form the {\em orthogonal group} $\ON$, i.e., the product of orthogonal matrices is also an orthogonal matrix~\cite{Gilmore:74:lgl}. Thus, we can have some control over complexity by defining transform matrices as the product of sparse orthogonal matrices.

We assume that, for each classification index $c$, we have a series of {\em transform parameter vectors} $\{\Bf{h}_{c,k}\}_{k=1}^{P_c}$, where $\Bf{h}_{c,k}$ has dimension $n_{c,k}$, and we also have $P_c$ mappings from the parameter vectors to orthogonal matrices
\begin{equation}
  \Bf{F}_{c,k} : \RealS^{n_{c,k}} \rightarrow \ON, \quad k = 1, 2, \ldots, P_c,
\end{equation}
and define the transform orthogonal matrix as the product
\begin{multline}
 \label{eq:CMPTprd}
 \Bf{T}_c(\Bf{h}_{c,1}, \cdots, \Bf{h}_{c,P_c}) = \\ 
  \Bf{F}_{c,P_c}(\Bf{h}_{c,P_c}) \cdots \Bf{F}_{c,2}(\Bf{h}_{c,2}) \Bf{F}_{c,1}(\Bf{h}_{c,1}).
\end{multline}
The inverse transform is
\begin{multline}
 \Brack{ \Bf{T}_c(\Bf{h}_{c,1}, \cdots, \Bf{h}_{c,P_c}) }^{-1} = \\ 
  \Brack{ \Bf{F}_{c,1}(\Bf{h}_{c,1}) }^{-1} \Brack{ \Bf{F}_{c,2}(\Bf{h}_{c,2}) }^{-1} \cdots \Brack{ \Bf{F}_{c,P_c}(\Bf{h}_{c,P_c}) }^{-1},
\end{multline}
i.e., the sequence of inverse transforms in reverse pass order.

Under these assumptions, the signal-adaptive transform that we propose, which we call {\em Parametric Multi-pass Transform} (PMPT), can be interpreted as having the residual transform being computed with a series of {\em transform passes.}

To be computationally efficient, we also require each PMPT pass to have O($N$) computational complexity, and thus we must have
\begin{Enumerate}
\item A pass orthogonal matrix $\Bf{F}_{c,k}$ is defined by a number of parameters ($n_{c,k}$) proportional to $N$;
\item The number of arithmetic and logical operations in each pass is proportional to $N$;
\item The pass operations can be parallelized and evenly distributed to $N/D$ processors, where $D$ is a small integer (e.g., 1, 2, 4).
\end{Enumerate}

This last requirement guarantees that the new transform can be computed most efficiently in both serial and parallel implementations. It also means that some known orthogonal matrix factorizations~\cite{Vaidyanathan:93:mfb,Strang:09:ila,Chen:12:dol} cannot be directly used for PMPT, since they cannot be efficiently computed in parallel. 

Under those assumptions, if we employ a constant number of passes $P$, then we obtain the computational complexity shown in the last row of Table~\ref{tb:ExpBlkCmplx}.

\section{Hypercube-Givens Transform (HyGT)} \label{sc:HyGT}

\begin{figure}
\centering
\includegraphics[width=60mm]{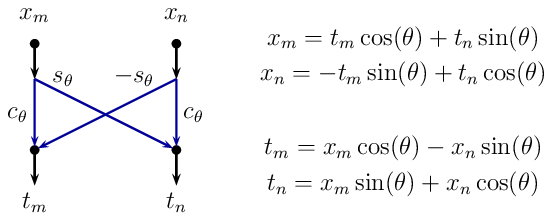}
\caption{\label{fg:HyGTBut}``Butterfly'' structure for computing one Givens rotation by angle $\theta$, on a 2-D subspace defined by indexes $m$ and $n$.}
\end{figure}

The special case of PMPT that we analyze here employs Givens rotations, commonly represented with ``butterflies'' as shown in Fig.~\ref{fg:HyGTBut}, as the basic structure for defining sparse parameterized orthogonal transforms.

A Givens rotation matrix $\Bf{G}(m,n,\theta)$, defined by~\cite{Vaidyanathan:93:mfb}
\begin{equation}
 G_{i,j}(m,n,\theta) = \begin{MValue}
   \cos(\theta), & i = j = m, \mbox{ or } i = j = n, \\
   \sin(\theta), & i = m, \; j = n, \\
   -\sin(\theta), & i = n, \; j = m, \\
   1, & i = j \mbox{ and } i \neq m \mbox{ and } i \neq n, \\
   0, & \mbox{otherwise,}
  \end{MValue} \nonumber
\end{equation}
rotates the two-dimensional vector of elements with indexes $m$ and $n$ by an angle $\theta$.

\begin{figure}
\centering
\includegraphics[width=88mm]{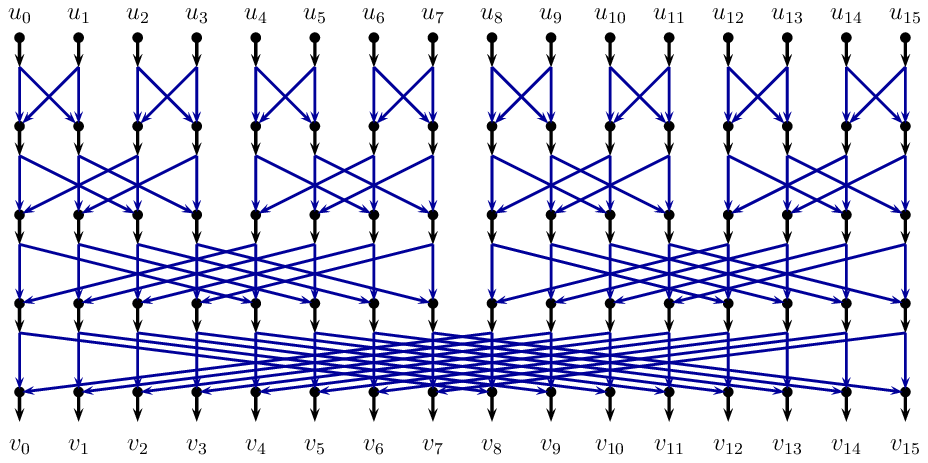}
\caption{\label{fg:HyGTPass}Each round of the Hypercube-Givens Transform (HyGT) corresponds to $\log_2(N)$ different PMPT passes. In this example, $N=16$, and we have 4 passes, each corresponding to a different dimension of a 4-dimensional hypercube.}
\end{figure}

Individual Givens rotation matrices are used in many applications, but for computationally efficient transforms we must use a form that enables parallel computations (similarly to techniques for using the Jacobi method in parallel processors~\cite{Brent:85:ssv}). Assuming that $N$ is even, we define vectors $\Bf{m}$, $\Bf{n}$, each with $N/2$ indexes such that
\begin{equation}
 \label{eq:PairCover}
 \bigcup_{k=0}^{N/2-1} \{ m_k, n_k \} = \{ 0, 1, 2, \ldots, N - 1 \}.
\end{equation}
and a vector $\boldsymbol{\theta}$ with $N/2$ angles, and define the {\em parallel-Givens} matrix as
\begin{equation}
 \label{eq:PGivens}
 \Bf{G}(\Bf{m}, \Bf{n}, \BS{\theta}) = \prod_{k=0}^{N/2-1} \Bf{G}(m_k,n_k,\theta_k).
\end{equation}
Note that here we can use this notation to define a matrix because, if eq.~(\ref{eq:PairCover}) is satisfied, then all the matrix multiplications in eq.~(\ref{eq:PGivens}) are commutative.

In addition, as shown in Fig.~\ref{fg:HyGTBut}, the inverse of a Givens rotation with angle $\theta$ is the same type of rotation, but with angle $-\theta$, and thus we have
\begin{equation}
 \label{eq:PGivensInv}
 \Brack{ \Bf{G}(\Bf{m}, \Bf{n}, \BS{\theta}) }^{-1} = \Bf{G}(\Bf{m}, \Bf{n}, -\BS{\theta}).
\end{equation}

The transforms defined by the parallel-Givens matrix already satisfy all the complexity requirements of a PMPT pass, listed in Section~\ref{sc:CMPT}. What we do next is to consider a particular case, called HyGT (Hypercube-Givens Transform, pronounced as ``height''), where we eliminate the need to have the index vectors $\Bf{m}$ and $\Bf{n}$ explicitly included as transform parameters.

This property is very important, since it is possible to have equally good transforms computed with a smaller number of ``butterflies,'' but that in the end use significanly more memory, due to the need for index storage.

Assuming that $N$ is a power of two, we define a HyGT {\em round} as a sequence of $\log_2(N)$ passes, where in each pass the indexes in vectors $\Bf{m}$ and $\Bf{n}$ are defined by edges of a hypercube with dimension $\log_2(N)$, sequentially in each direction. This may appear complicated, but the hypercube arrangement is quite common, and found, for instance, in descriptions of the Fast-Fourier-Transform (FFT) algorithm~\cite{Oppenheim:09:dsp}. Fig.~\ref{fg:Code} shows the C++ function to efficiently compute arrays of indexes.
 
\begin{figure}
\centering
\includegraphics[width=72mm]{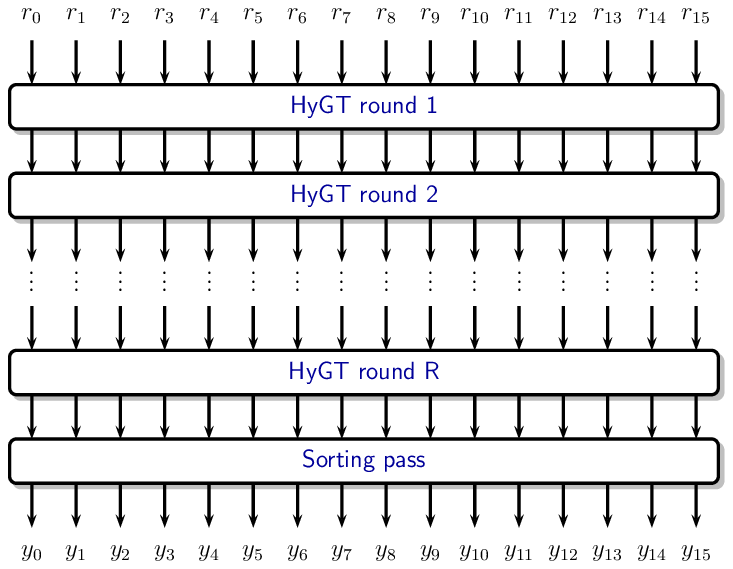}
\caption{\label{fg:HyGTRounds}The complete HyGT is composed of one or more rounds, each with $\log_2(N)$ passes, and an optional final sorting (permutation) pass.}
\end{figure}

\begin{figure}
\begin{tt}\begin{footnotesize}\begin{verbatim}
void Hypercube (int log2_N, int * m[], int * n[])
{
  for (int i = 0; i < log2_N; i++) {
    int hN = 1 << (log2_N - 1), k = 1 << i;
    for (int j = 0; j < hN; j++) {
      m[i][j] = j + (j & -k);
      n[i][j] = m[i][j] + k; } }
}
\end{verbatim}\end{footnotesize}\end{tt}\vspace{-3ex}
\caption{\label{fg:Code}C++ code to compute arrays of hypercube indexes.}
\end{figure}

In practice we need more than one HyGT round to obtain good compression. As shown in Fig.~\ref {fg:HyGTRounds}, the full transform is composed of $R$ rounds, and may include an optional permutation pass, to sort transform coefficients according to their variance. The inverse HyGT is implemented using~(\ref{eq:PGivensInv}), i.e., using negative angles, and computing the passes in reverse order.

Note that the angle used in each butterfly is a different parameter, and thus the total number of transform parameters is $R N \log_2(N) / 2$ angles (which correspond to the parameter vectors $\Bf{h}_{c,k}$ of Section~\ref{sc:CMPT}). This is quite different from other transforms, and from other forms of factorization using Givens rotations~\cite{Vaidyanathan:93:mfb,Strang:09:ila}.

\begin{figure}
\centering
\includegraphics[width=86mm]{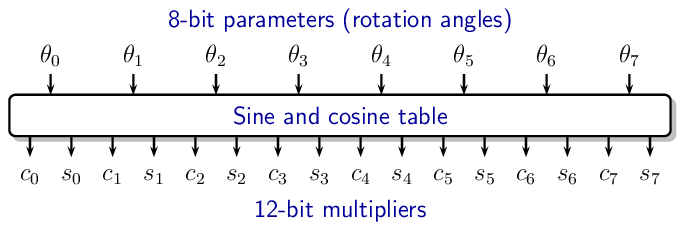}
\caption{\label{fg:SCTable}System for ``on-the-fly'' conversion of transform parameters.}
\end{figure}

One additional advantage of the factorization using Givens rotations is that it allows for a simple method to further reduce memory requirements, without loss in compression. As shown in Fig.~\ref{fg:SCTable}, we can use a system where the transform parameters are represented with a small number of bits, and are converted to higher precision multipliers only when needed. The table with sine and cosine values does not add significantly to memory usage, since it is shared by all transforms.

Note that this property is quite important for video coding standards, since their transforms must be implemented with well-defined integer arithmetic. As shown in the simulation results of Section~\ref{sc:ExpRes}, HyGT can approximate video coding KLTs using only one byte per parameter.

\section{HyGT-parameter optimization}
\label{sc:HyGTOpt}

Transforms should ideally be optimized by measuring how they affect the average rate-distortion performance of a given encoder. However, this requires extremely long sets of simulations, and it is normally used only for final fine-tuning. The practical approach is to use approximations, and a simplified model of the encoder, using only the variances of the transformed residuals for rate and distortion estimations.

To simplify the notation, here we drop subscripts $c$ indicating the dependence on the classification index, use a single vector $\Bf{h}$ to represent all the HyGT parameters combined, and represent the transform by the $N \times N$ matrix function $\Bf{T}(\Bf{h})$.

Using this notation, we define the vector with the variances of transformed residuals as
\begin{eqnarray}
 \Bf{v}(\Bf{h}) & = & \mbox{diag}(\Bf{T}(\Bf{h}) \, \BM{\Phi} \, \Trns{T}(\Bf{h}) ) \\
   & = & [ \begin{array}{cccc}
    \sigma_0^2(\Bf{h}) & \sigma_1^2(\Bf{h}) & \cdots & \sigma_{N-1}^2(\Bf{h}) \end{array} ]^{\mbox{\sf \tiny T}}, \nonumber
\end{eqnarray}
where the correlation matrix $\BM{\Phi}$ is defined by eq.~(\ref{eq:CorrDefn}).

It is common to measure coding gains using the high-rate assumption, and the following approximations for squared-error distortion $D$, and bit rate $R$ (in bits per sample)~\cite{Woods:11:msp,Taubman:02:jsi,Pearlman:11:dsc}
\begin{equation}
 \label{eq:HRRD}
 D(R) \approx \varepsilon^2 \sigma^2 \, 2^{-2R}, \qquad
 R(D) \approx \Half \Funcf{\log_2}{\varepsilon^2 \sigma^2}{D},
\end{equation}
where $\sigma^2$ is the variance, and $\varepsilon$ is a constant that depends on the distribution (Gaussian, Laplacian etc.).

For optimization of HyGT parameters (vector $\Bf{h}$), we use a variation of~(\ref{eq:HRRD}) based on Laplacian distribution to approximate rate-distortion, and a non-linear optimization method that searches for $\Bf{h}$ that produces the highest transform coding gain. Since the corresponding optimization problem is non-convex, we repeat the search with different initializations and pick the best local optimal solution. Further details of the optimization are out of the scope of this paper.

\section{Experimental results}\label{sc:ExpRes}

\begin{table*}
\centering
\caption{\label{tb:FullNumRes} BD-rate results for video coding simulations (all intra) using KLT and HyGT as secondary transforms, memory usage ratio 
for storing sets of 105~transforms per block size.} 
\renewcommand{\arraystretch}{1}
\begin{tabular}{|c|l||c||c|c||c|c|c||c|c|} \hline

 \multirow{2}{*}{\bf Resolution} & \multirow{2}{*}{\bf Sequence} & \multicolumn{8}{|c|}{\rule{0pt}{2.4ex} \bf Coding Scheme} \\ \cline{3-10}
 & & \rule{0pt}{2.4ex} K/K & H(2)/K & H(3)/K  & K/H(3) & K/H(4) & K/H(5) & H(2)/H(3) & H(2)/H(4)  \\ \hline \hline
 
 \multirow{8}{*}{4K} 
 & Tango
 &     -8.0\% &     -8.0\% &     -8.0\% &     -8.4\% &     -8.6\% &     -8.7\% &     -8.4\% &     -8.6\% \\
 & Drums
 &     -5.0\% &     -5.0\% &     -5.0\% &     -5.5\% &     -5.6\% &     -5.6\% &     -5.5\% &     -5.6\% \\
 & CampfireParty
 &     -5.5\% &     -5.4\% &     -5.5\% &     -5.4\% &     -5.5\% &     -5.5\% &     -5.4\% &     -5.5\% \\
 & ToddlerFountain
 &     -4.5\% &     -4.5\% &     -4.5\% &     -4.8\% &     -4.8\% &     -4.8\% &     -4.8\% &     -4.8\% \\ \cline{2-10}
 & CatRobot
 &     -6.4\% &     -6.4\% &     -6.4\% &     -6.5\% &     -6.6\% &     -6.6\% &     -6.5\% &     -6.6\% \\
 & TrafficFlow
 &    -12.3\% &    -12.3\% &    -12.3\% &    -11.9\% &    -12.3\% &    -12.4\% &    -11.9\% &    -12.3\% \\
 & DaylightRoad
 &     -5.9\% &     -5.9\% &     -5.9\% &     -6.1\% &     -6.1\% &     -6.1\% &     -6.1\% &     -6.1\% \\
 & Rollercoaster
 &    -10.0\% &     -9.9\% &    -10.0\% &    -10.1\% &    -10.4\% &    -10.5\% &    -10.1\% &    -10.4\% \\ \hline
 \multirow{5}{*}{1920 $\times$ 1080}
 & Kimono
 &     -4.8\% &     -4.7\% &     -4.7\% &     -5.3\% &     -5.3\% &     -5.4\% &     -5.3\% &     -5.3\% \\
 & ParkScene
 &     -4.6\% &     -4.7\% &     -4.6\% &     -4.7\% &     -4.8\% &     -4.8\% &     -4.7\% &     -4.8\% \\
 & Cactus
 &     -5.4\% &     -5.4\% &     -5.4\% &     -5.5\% &     -5.6\% &     -5.7\% &     -5.6\% &     -5.6\% \\
 & BasketballDrive
 &     -4.2\% &     -4.1\% &     -4.1\% &     -4.3\% &     -4.3\% &     -4.3\% &     -4.3\% &     -4.3\% \\
 & BQTerrace
 &     -4.9\% &     -4.9\% &     -4.9\% &     -5.0\% &     -5.1\% &     -5.1\% &     -5.0\% &     -5.1\% \\ \hline
 \multirow{4}{*}{832 $\times$ 480}
 & BasketballDrill
 &    -11.6\% &    -11.5\% &    -11.6\% &    -11.1\% &    -11.5\% &    -11.6\% &    -11.1\% &    -11.4\% \\
 & BQMall
 &     -3.9\% &     -3.9\% &     -3.9\% &     -4.1\% &     -4.2\% &     -4.1\% &     -4.2\% &     -4.2\% \\
 & PartyScene
 &     -3.8\% &     -3.8\% &     -3.8\% &     -3.8\% &     -3.8\% &     -3.8\% &     -3.8\% &     -3.8\% \\
 & RaceHorses
 &     -5.5\% &     -5.5\% &     -5.6\% &     -5.5\% &     -5.6\% &     -5.7\% &     -5.5\% &     -5.6\% \\ \hline
 \multirow{3}{*}{416 $\times$ 240}
 & BasketballPass
 &     -4.6\% &     -4.6\% &     -4.7\% &     -4.8\% &     -4.7\% &     -4.8\% &     -4.7\% &     -4.8\% \\
 & BQSquare
 &     -4.1\% &     -4.1\% &     -4.2\% &     -4.1\% &     -4.1\% &     -4.2\% &     -4.2\% &     -4.2\% \\
 & BlowingBubbles 
 &     -5.0\% &     -5.0\% &     -4.9\% &     -4.9\% &     -5.0\% &     -5.0\% &     -4.9\% &     -5.0\% \\
 & RaceHorses
 &     -6.6\% &     -6.6\% &     -6.6\% &     -6.5\% &     -6.6\% &     -6.7\% &     -6.4\% &     -6.6\% \\ \hline
 \multirow{3}{*}{1280 $\times$ 720}
 & FourPeople
 &     -6.1\% &     -6.2\% &     -6.2\% &     -6.2\% &     -6.3\% &     -6.3\% &     -6.3\% &     -6.3\% \\
 & Johnny
 &     -5.2\% &     -5.3\% &     -5.2\% &     -5.4\% &     -5.4\% &     -5.5\% &     -5.5\% &     -5.5\% \\
 & KristenAndSara
 &     -6.2\% &     -6.3\% &     -6.2\% &     -6.3\% &     -6.4\% &     -6.4\% &     -6.3\% &     -6.4\% \\ \hline \hline
\multicolumn{2}{|c||}{\rule{0pt}{2.2ex} \bf BD-rate average}
 & \bf -6.0\% & \bf -6.0\% & \bf -6.0\% & \bf -6.1\% & \bf -6.2\% & \bf -6.2\% & \bf -6.1\% & \bf -6.2\% \\ \hline \hline
\multicolumn{2}{|c||}{Memory usage ratio, 4$\times$4}
 &     1.0 &     4.0 &     2.7 &     1.0 &     1.0 &     1.0 &     4.0 &     4.0 \\
\multicolumn{2}{|c||}{Memory usage ratio, 8$\times$8}
 &     1.0 &     1.0 &     1.0 &     7.1 &     5.3 &     4.3 &     7.1 &     5.3 \\
\multicolumn{2}{|c||}{\bf Memory usage ratio, average}
 & \bf 1.0 & \bf 1.0 & \bf 1.0 & \bf 5.2 & \bf 4.3 & \bf 3.6 & \bf 6.8 & \bf 5.2 \\ \hline
\end{tabular}
\end{table*}

In this section, we compare (i) coding performance and (ii) memory requirement, between HyGTs and KLTs. The coding performance is measured in terms of BD-rate \cite{Bjontegaard:01:cap} obtained by coding several standard video sequences under the {\em All-Intra} coding configuration, as specified in the common test conditions (CTC)~\cite{Wien:15:HEV,Bjontegaard:01:cap,Bossen:11:ctc} and benchmarking against the reference software (HM~16.6) of the state-of-the-art video coding standard (HEVC).\footnote{https://hevc.hhi.fraunhofer.de/svn/svn\_HEVCSoftware/}
In our experiments, both HyGTs and KLTs are used as secondary transforms, which are non-separable transforms applied on a sub-block of low-frequency coefficients obtained by the (primary) separable transform chosen by the encoder~\cite{Zhao:15:nss,Zhao:16:ins}.\footnote{Further details are available in reference~\cite{Zhao:15:nss,Zhao:16:ins} proposed for possible inclusion in a future ITU/MPEG video compression standard.} In order to control the complexity for large blocks, we only design 4$\times$4 and 8$\times$8 block transforms (i.e., HyGTs and KLTs), where 
\begin{itemize}
\item for 4$\times$4 blocks, we use a 4$\times$4 HyGT/KLT;
\item for 8$\times$8 or larger blocks, we use a 8$\times$8 HyGT/KLT applied to top-left sub-block of transform coefficients. 
\end{itemize}
Note that both HyGTs and KLTs are designed by training on a dataset of transform coefficients generated using the JEM-2.0 video coding software.\footnote{https://jvet.hhi.fraunhofer.de/svn/svn\_HMJEMSoftware/tags/HM-16.6-JEM-2.0} The same software is used to test the performance of designed transforms on a different video dataset given in the common test conditions (CTCs). To achieve better compression, three different HyGTs/KLTs are designed for 4$\times$4 and 8$\times$8 secondary transforms for each of 35~intra modes, so that the encoder has flexibility of choosing, in rate-distortion sense, the best transform out of three transform candidates, or the option of not applying a secondary transform. This requires storing total of 210 different transforms at both encoder and decoder.

The Table \ref{tb:FullNumRes} compares (i)~BD-rate \cite{Bjontegaard:01:cap} and (ii)~memory requirements, which is in terms of
\begin{equation}
\text{memory usage ratio} = \frac{\text{memory used by KLTs}}{\text{memory used by HyGTs}} ,
\end{equation}
for different transform coding schemes. The columns of the table include the results obtained by applying different pairs of transforms applied on 4$\times$4 and 8$\times$8 blocks, where A/B indicates that transforms A and B are applied to 4$\times$4 and 8$\times$8 blocks, respectively. Moreover in Table~\ref{tb:FullNumRes}, K denotes the KLT and H($R$) represents the HyGT with $R$ rounds. The results show that HyGTs can significantly reduce memory requirement while achieving similar compression performance (6\% coding gain) compared to KLT-based transform coding. 

\section{Conclusions}
\label{sc:Concl}

In this work, we have proposed a novel method allowing us to design signal-adaptive transforms that provide high compression efficiency with low computational complexity. The resulting transforms (HyGTs) lead to a special type of sparse orthogonal matrices, which can be efficiently parallelized and require significantly less memory. Our extensive experimental results have shown that the proposed HyGTs closely approach to the coding gain achived by traditional KLT-based solutions ($\sim$6\%), while significantly reducing the memory requirement ($\sim$5-7 times less memory).


\end{document}